\documentclass{neuthist18}

\def\be{\begin{equation}}
\def\ee{\end{equation}}
\def\bea{\begin{eqnarray}}
\def\eea{\end{eqnarray}}

\newcommand{\Photo}{}

\begin{document}
\vspace*{4cm}
\title{A Brief History of the Co-evolution of Supernova Theory with Neutrino Physics}

\author{Adam Burrows}

\address{Department of Astrophysical Sciences, Princeton University,\\ Princeton, NJ USA 08544}

\maketitle\abstracts{
 The histories of core-collapse supernova theory and of neutrino physics have paralleled one 
another for more than seventy years.  Almost every development in neutrino physics necessitated
modifications in supernova models.  What has emerged is a complex and rich
dynamical scenario for stellar death that is being progressively better tested by increasingly 
sophisiticated computer simulations. Though there is still much to learn about the agency and
details of supernova explosions, whatever final theory emerges will have the 
neutrino at its core. I summarize in this brief contribution some of the salient developments
in neutrino physics as they related to supernova theory, while avoiding any attempt 
to review the hundreds of pivotal papers that have pushed supernova theory forward.
My goal has been merely to highlight the debt of supernova astrophysics to neutrino physics.}

\section{Introduction}
\label{intro}

The theory of the violent deaths of massive stars in what are called supernova explosions
has a long pedigree that spans more than half a century, has engaged hundreds of 
researchers, and has proven more elusive than anticipated.  However, with the advent of numerically
and physically sophisticated codes with which to simulate the onset of explosion in three
spatial dimensions, the theoretical community now seems to be zeroing in on the 
mechanism of explosion.  Central to this emerging theory are the neutrinos 
of all species produced copiously at the high densities and temperatures achieved 
during and after the collapse of the unstable Chandrasekhar core created in the center
of the massive star at the end of its life. A fraction of these emerging neutrinos are absorbed
behind the bounce shock wave to drive it into explosion, aided by the turbulence in the outer core
driven predominantly by neutrino heating itself.  In this context, the interaction of neutrinos 
with matter is key to the fidelity with which theorists can simulate the explosion phenomenon.
Hence, progress in supernova theory has paralleled developments not only in computational
capabilities, but also in advances in our understanding of neutrino physics. 
Developments in nuclear physics over the years have also played (and continue to 
play) important roles, as have improvements in our understanding of pre-supernova stellar evolution.
Nevertheless, if neutrinos prove to be the agents of explosion then their pivotal role
deserves to be highlighted. This parallel evolution over the years in our understanding 
of neutrinos and of supernovae is the subject of this brief paper.  I will focus upon rough 
timelines for important conceptual progress in both spheres and will omit almost any detailed 
discussion of supernova theory itself.  Such can be found in numerous papers in the archival 
literature \cite{janka2012,burrows2013,lentz:15,burrows:18,oconnor_couch2018,TaKoSu16,vartanyan2019}.

\section{The 1930s}
\label{1930}

Of course, the 1930s were the decade of the discovery of the neutron, the postulation 
of the existence of the neutrino by Wolfgang Pauli, and its naming by Enrico Fermi.  
Fermi also formulated the point interaction to describe $\beta$-decay.  In addition, this decade witnessed 
the publication of the prescient papers by Baade \& Zwicky \cite{1934PNAS...20..254B,1934PNAS...20..259B},
wherein they postulated the existence of neutron stars, coined the term ``supernova," and
connected supernovae with the creation of both neutron stars and cosmic rays.  However, the possible 
connection between supernova and neutrinos was not in the air.

\section{The 1940s}
\label{1940}

This connection was first made by Gamow \& Schoenberg \cite{1941PhRv...59..539G} in their April 1, 1941
paper. They noted that stellar interiors could be hot and could emit neutrinos that might 
accelerate evolution by facilitating rapid core contraction.  This contraction was to be
accompanied by rapid outward motion of the envelope and an explosion, and these
explosions could be either novae or supernovae, depending upon the original stellar mass.
However, the neutrino emission mechanism was by a cycle of electron capture and subsequent
beta decay, the so-called ``URCA" process (invented in this paper), which did not result in 
a net change in composition and is not relevant in the modern contexts,
but did result in the loss of energy by volumetric neutrino pair emission. The precise mechanism
by which the outer layers were to be expelled was not explained, and it was certainly not by neutrino
energy or momentum deposition. Neutrinos did not directly drive the explosion, but their emission
was to lead, by a mechanism unexplained, to a dynamical stellar phase. Aside from its intriguing comingling of supernovae 
with neutrinos, this paper was too short on details and too far from what we now conclude concerning
stars, supernovae, novae, and neutrinos to be considered useful.  In fact, there followed
a gap of $\sim$20 years before the neutrino-supernova saga would reemerge. And as far as neutrino
physics itself was concerned, the world in the 1940s was too engaged in other pursuits to generate much
of substance in the open literature.

\section{The 1950s}
\label{1950}

After the World War there was a great deal of progress in particle physics, nuclear physics, 
and (not unexpectedly) the theory of nucleosynthesis.  The latter is exemplified 
by the publications of Burbidge, Burbidge, Fower, \& Hoyle \cite{1957RvMP...29..547B} and 
Cameron \cite{1957PASP...69..201C}. These papers represented a growing literature in which 
great strides were made in understanding the origin of the elements in stars, either during quiescent 
burning or explosively.  In physics, the weak interaction experienced a great deal 
of ferment. Lee \& Yang \cite{Lee:1956qn} suggested parity violation in 
the weak interaction and Wu \cite{Wu:1957my} demonstrated it.  Goldhaber et al. 
\cite{Goldhaber:1958nb} explored helicity, Pontecorvo \cite{Pontecorvo:1957qd}
posited $e^-/\mu^-$ universality and neutrino oscillations, and Feymann 
\& Gel'Mann \cite{Feynman:1958ty} introduced $V-A$ theory.  Notably, Cowan \& Rienes \cite{Reines:1953pu}
actually detected and measured (anti-)neutrinos from a reactor.  However, aside from ongoing speculation, 
there was not much progress in the theory of supernova explosions.

\section{The 1960s}
\label{1960}

This changed with the study by H.Y. Chiu of thermal neutrino emission processes in stars
\cite{1961PhRv..123.1040C,1961PhRv..122.1317C,1964AnPhy..26..364C}.  Relevant in
there own right, these studies also led to discussions between Chiu and Stirling Colgate,
which, along with the latter's participation in the Partial Test Ban Treaty 
negotiations, inspired Colgate to take the next big step in supernova 
theory (S. Colgate, private communication).  In 1966, Colgate \& White 
\cite{1966ApJ...143..626C} published the notion that copious neutrino 
production in supernova progenitor cores immediately after core collapse could, 
through the agency of neutrino heating, unbind the mantle in a supernova explosion.  
In this paradigm, the supernova shock wave would be driven by neutrino energy deposition, 
and this is, in very broad outline and with a few remaining caveats, the currently 
accepted mechanism. In 1967, Dave Arnett \cite{1967CaJPh..45.1621A} conducted a 
numerically and physically more sophisticated ``radiation/hydrodynamic" study, also including 
muon neutrinos, and the stage was set for the permanent association between neutrinos 
and core-collpase supernova explosions.  On the physics front, Lederman et al. \cite{Danby:1962nd}
detected the $\nu_{\mu}$ neutrino and the Weinberg/Salam/Glashow theory of electroweak unification
was formulated. The latter postulated the existence of neutral currents.

\section{The 1970s}
\label{1970}

This decade provided most of the remaining progress in the understanding 
of the weak interaction, neutrinos, and the neutrino-matter interaction
necessary to simulate supernova with modern physical fidelity. The weak neutral
current, theorized in the last decade, was measured early in this. This led
to the calculation by Dicus \cite{1972PhRvD...6..941D} of neutral-current 
scattering of neutrinos off free nucleons, forbidden in the $V-A$ theory
of Feynmann and Gel'Mann, as well as the calculation by Freedman \cite{Freedman:1973yd} of 
neutral-current scattering off nuclei. Both these processes can dominate
the neutrino-matter scattering rate at various phases of core collapse
and explosion and were unknown to earlier supernova theory. In addition,
M. Perl and collaborators discovered the $\nu_{\tau}$ neutrino and 
Wolfenstein \cite{Wolfenstein:1977ue} introduced matter effects into 
neutrino oscillation theory. 

On the supernova front, one of the most important developments was the realization that 
upon core collapse, at the progressively higher densities and temperatures achieved, 
not only does the optical depth to neutrinos become large and does neutrino diffusion 
from the ``proto-neutron star" \cite{1981ApJ...251..325B} become relevant, but that electron 
lepton number becomes trapped. Trapping is not only the achievement of high neutrino optical depths, 
but the cessation on the dynamical timescales of core collapse and infall of the net loss of electron 
lepton number ($\rm{Y}_e$ = electron/baryon ratio). Rather than achieving low neutron-star 
electron numbers of $\sim$0.03 before reaching nuclear densities, electron fractions near $\sim$0.3 are 
frozen in.  This is due to the onset during collapse of the inverse reaction of $\nu_e$ capture
onto protons in and out of nuclei that pushes the matter into a ``$\beta$"(chemical)-equilibrium 
and preserves the electrons. The trapped electron neutrinos are then further compressed 
during the later stages of collapse, but they are now degenerate fermions. Trapping was first
recognized by Ted Mazurek \cite{1974Natur.252..287M} using the old weak-interaction theory
and by Kats Sato \cite{sato} using the new theory with neutral currents. 

Further compression elevates the electron neutrino chemical potential ($\sim$Fermi energy), and the average  $\nu_e$
neutrino energy at and subsequent to bounce soars to $\sim$150-300 MeV at the center. Due to 
the stiffly increasing interaction cross sections with increasing neutrino energy, the optical
depth to $\nu_e$ diffusion grows to $\sim$$10^5$.  Such a large optical depth translates 
into a time of many seconds for the diffusion of lepton number and energy out of the 
proto-neutron star (PNS) \cite{burlat86}. Hence, trapping leads directly to the long emission 
times of supernova neutrinos.  Without the recognition of neutrino trapping, the duration 
of a supernova neutrino burst would have been predicted to be less than $\sim$100 milliseconds.

The final overarching physical piece of the collapse puzzle was the recognition that 
the collapsing core was a Chandrasekhar-mass white dwarf supported by electron
degeneracy.  This was not obvious until the requisite thermal neutrino heating in the 
stellar core was implemented in stellar evolution codes and the thermostatic effects of
the excited states of nuclei were incorporated into the equation of state.  The upshot
was the lower entropies and lower temperatures that ensured the core pressures were 
due to electron degeneracy pressure from the onset of and during core collapse 
to nuclear densities.

\section{The 1980s}
\label{1980}

The 1980s saw the discovery of the $W$ and $Z$ bosons by the teams led by Rubbia and 
van de Meer \cite{Arnison:1983mk} and the solidification of the electroweak theory. There were also important
developments in the solar neutrino puzzle and the recognition of the reality of neutrino 
oscillations.  On the supernova front, the physics of the neutrino-matter interaction 
was mature, as reflected in the summary paper by Steve Bruenn \cite{1985ApJS...58..771B}.
However, with sophistication in the neutrino sector came puzzles in supernova theory.
In particular, spherical models did not explode directly.  Jim Wilson broke the logjam with the
``delayed" neutrino heating mechanism, wherein the bounce shock stalled for hundreds 
of milliseconds, only to be revived thereafter.  He traced success to the boosting of
the driving neutrino luminosities after bounce by ``neutron-finger" convection in the 
inner core.  In his calculations, performed using enforced mixing-length convection,
the boost was $\sim$25\%, and this was enough to revitalize the explosion. However,
such convection was later shown by Bruenn \& Dineva \cite{bruenn_dineva} to be unphysical.
Nevertheless, the potential role of hydrodynamic instabilities and turbulence just interior
to the stalled shock wave, driven mostly by neutrino heating itself (from below) was found
to be crucial and this basic idea was later developed by among others Herant et al.\cite{herant1994} 
and Burrows et al.\cite{burrows:95}.  Such turbulence is now a central, perhaps enabling, 
facet of supernova theory.  However, the manifest hydrodynamic instabilities and turbulence
in supernova cores required the development of sophisticated multi-dimensional radiation/hydrodynamic
codes to simulate the supernova dynamics in its full multi-dimensional richness.
It is reasonably concluded that the development of such complicated codes and the expensive 
computational platforms they require has set the long timescale of subsequent progress up 
to the present day.  

However, the most exciting event in this decade at the interface between neutrinos and supernovae 
was the detection in 1987 by Hirata et al. \cite{Hirata:1987hu} in Kamiokande II and by 
Bionta et al. \cite{1987PhRvL..58.1494B} in the IMB of the neutrinos from the supernova
SN1987A in the Large Magellanic Cloud $\sim$50 kiloparsecs away.  This first, and to date only, such
detection galvanized the astrophysics and physics communities, generated hundreds of papers,
and established unambiguously the neutrino/supernova connection. The many-second 
duration of the event, with average event energies near $\sim$15 MeV, confirmed 
that 1) neutrinos are generated and radiated in abundance in supernova cores,
2) neutrinos diffuse out of the dense PNS, 3) the scale of the radiated energy is the binding energy 
of a neutron star ($\sim$$3\times 10^{53}$ ergs), and 4) electron lepton number is trapped. The latter
seems compelling since trapping theory converted a $\sim$50-100 millisecond event into a 
multi-second event, a duration that was predicted before 1987. However, the effect of 
neutrino oscillations on the detected signal still remains to be determined \cite{shaquann2018}.

\begin{figure}[p]
\includegraphics[width=1.0\textwidth, angle = 0]{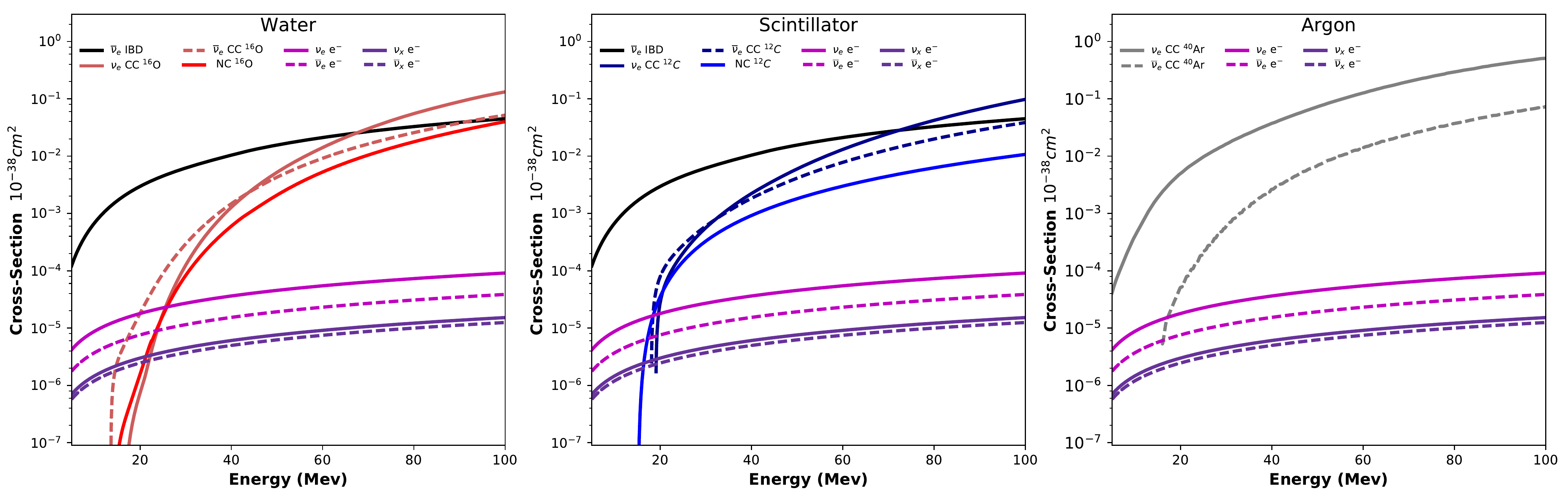}
\caption[]{Above are the energy-dependent cross sections for the neutrino-matter interactions in water (left),
scintillator (center), and liquid argon (right), taken from Seadrow et al. \cite{shaquann2018}. The water 
cross sections are relevant to water-Cherenkov detectors such as Super-K and Hyper-K.  The scintillator 
cross sections are relevant to such detectors as JUNO and the Argon cross sections are relevant to DUNE.
These cross sections were provided by the SNOwGLoBES software \cite{Scholberg:2012id}.}\label{cross}
\end{figure}

This, and witnessing supernova dynamics in real time, motivates the development of modern
supernova neutrino detection capabilities \cite{Scholberg:2012id}. It is only by capturing 
supernova neutrinos and the gravitational waves also generated during collapse that we can see what happens
at the time it is happening.  Otherwise, the core is shrouded in mystery by the profound opacity
to photons of the stellar envelope that surrounds it.  Currently, 
Super-Kamiokande \cite{Ikeda:2007sa}, IceCube \cite{abbasi:2011}, and various 
underground detectors in the Gran Sasso tunnel in Italy stand guard in 
anticipation of a galactic event, but in the near future JUNO \cite{An:2015jdp} 
and DUNE \cite{ankowski:2016} will join them and Hyper-K \cite{abe11} will replace 
Super-Kamiokande.  The per-particle interaction cross sections of importance in these detectors are
plotted in Figure \ref{cross}.  For each, it is clear from these plots which reactions 
dominate.  It is worth noting that these modern sentinels could capture hundreds
to many thousands of events from a galactic supernova at $\sim$10 kiloparsecs, whereas 
we culled but 11 (Kamioka II) $ and $ 8 (IMB) events from SN1987A.  Clearly, much remains to be learned.

\section{The Present}
\label{present}

The theory of core-collapse supernova has experienced significant development over the last few decades.
This progress has relied upon knowledge of the interaction of neutrinos with matter via production,
absorption, and scattering.  A set of important processes now incorporated into modern 
supernova codes is given in Table \ref{table1}. With the knowledge represented, theorists
have created sophisiticated computational capabilities that have enabled the exploration of 
the supernova mechanism and dynamics in its full multi-dimensional complexity.  The 
neutrino-driven mechanism in its basic form, despite a great deal of change over the decades
in our understanding of neutrinos and despite the necessary increase in theoretical sophistication, 
still holds pride of place $-$ published exploding models in three spatial dimensions are becoming common
\cite{vartanyan2019,lentz:15,melson:15b,roberts:16,TaKoSu16,muller2017,ott2018_rel,summa2018,oconnor_couch2018}.
Hence, though there is still much to resolve, the centrality of the neutrino in
this important astronomical context is assured. Given its weak coupling, modest beginnings in
theory, and multi-decade history, one may view its emergence as a pivotal player in one
of Nature's most violent natural phenomena as somewhat of a surprise.

\section*{Acknowledgments}

The author acknowledges support from the U.S. Department of Energy Office of Science and the Office
of Advanced Scientific Computing Research via the Scientific Discovery
through Advanced Computing (SciDAC4) program and Grant DE-SC0018297
(subaward 00009650). In addition, he gratefully acknowledges support
from the U.S. NSF under Grants AST-1714267 and PHY-1144374 (the latter
via the Max-Planck/Princeton Center (MPPC) for Plasma Physics).

\section*{References}

\bibliographystyle{unsrt}    
\bibliography{aburrows_sn}



\begin{table}[p]
\caption[]{Neutrino-matter reactions of primary relevance in the
core-collapse supernova context. BRT refers to Burrows, Reddy, \& Thompson\cite{burrows:06}
and BT refers to Burrows \& Thompson\cite{burrows_thompson2004}, which contains detailed
discussions of the handling of inelasticity for both neutrino-electron and
neutrino-nucleon scattering and of one approach to nucleon-nucleon bremsstrahlung.
TBH refers to Thompson, Burrows, \& Horvath\cite{thomp_bur_horvath}, where a detailed derivation
of the nucleon-nucleon bremmstrahlung rates can be found.
BS98 refers to Burrows \& Sawyer\cite{1998PhRvC..58..554B}, which contains the non-relativistic
dynamic structure factor formalism that informs an approach to
neutrino-nucleon inelastic scattering. R99 refers to Reddy et al.\cite{reddy1999},
where the relativistic formalism for inelasticity in neutrino-electron
scattering is provided. H02 refers to Horowitz\cite{horowitz2002}, where
corrections for weak magnetism are to be found.}
\label{table1}
\vspace{0.4cm}
\begin{center}
\begin{tabular}{lrcll}
\hline
\( \nu_i + A \)&
\( \rightleftharpoons  \)&
 \( \nu_i + A \)&
BRT\\
\( \nu_i + \mathrm{n,p} \)&
\( \rightleftharpoons  \)&
 \( \nu_i + \mathrm{n,p} \)&
BRT; BT, BS98\\
\( \nu _{\mathrm{e}} + \mathrm{n} \)&
\( \rightleftharpoons  \)&
 \( \mathrm{e}^{-} + \mathrm{p} \)&
BRT; H02\\
\( \bar{\nu }_{\mathrm{e}} + \mathrm{p} \)&
\( \rightleftharpoons  \)&
 \( \mathrm{e}^{+} + \mathrm{n} \)&
BRT; H02\\
\( \nu _{\mathrm{e}} + A' \)&
\( \rightleftharpoons  \)&
 \( \mathrm{e}^{-} + A \)&
Bruenn\cite{1985ApJS...58..771B}\\
\( \nu_i + \bar{\nu }_i \)&
\( \rightleftharpoons  \)&
 \( \mathrm{e}^{-} + \mathrm{e}^{+} \)&
BRT \\
\( \nu_i + \mathrm{e}^{-} \)&
\( \rightleftharpoons  \)&
 \( \nu_i + \mathrm{e}^{-} \)&
BRT; BT, R99\\
\( (\mathrm{n,p}) + (\mathrm{n,p}) \)&
\( \rightleftharpoons  \)&
 \( (\mathrm{n,p}) + (\mathrm{n,p}) + \nu_i \bar{\nu}_i \)&
BRT; BT, TBH\\
\hline
\end{tabular}
\end{center}
\end{table}

\end{document}